# Three-Leaf Quantum Interference Clovers in a Trigonal Single-Molecule Magnet

James H. Atkinson[1], Ross Inglis[2], Enrique del Barco[1*] and Euan K. Brechin[2*]

[1] *Department of Physics, University of Central Florida, Orlando, Florida 32765, USA*

[2] *EaStCHEM School of Chemistry, The University of Edinburgh, West Mains Road, Edinburgh, EH9 3JJ, UK*

*To whom correspondence should be addressed: E-mails: delbarco@physics.ucf.edu,

**The control of spin-orbit coupling (SOC) at the microscopic level has become a major research quest within the magnetism and spintronics communities. Altering the interaction between the spin and orbital degrees of freedom can fundamentally alter the electronic structure of a system, including electronic band inversion, and lead to new states of matter such as topological insulators. Whether based on high SOC lanthanides or transition metal ions, where the crystal field often quenches the orbital angular momentum, single-molecule magnets are an excellent playground for the study of the more fundamental aspects of this interaction and its intrinsic symmetries. This is the case of the molecular magnet reported in this article, where the trigonal symmetry imposed by the spatial arrangement of three constituent manganese ions and the corresponding orientations of their SOC anisotropy tensors result in a fascinating three-fold angular modulation of the quantum tunneling of the magnetization (QTM) rates never before observed, as well as in trigonal quantum interference patterns that mimic the form of a three-leaf clover. Interestingly, although expected in all the QTM resonances for a trigonal molecular**

**symmetry, the three-fold modulation only appears at resonances for which a longitudinal magnetic field is applied (i.e. resonances numbers |k| > 0). At k = 0, where no longitudinal field is present, the QTM probability displays a six-fold transverse field modulation. This comes as a direct consequence of a three-fold corrugation of the hard anisotropy plane, a predicted but previously unobserved feature which acts as an effective internal longitudinal field that varies the precise conditions required to maintain a resonance when a transverse field is applied. The sophisticated behavior of the QTM in this molecule allows an unequivocal association of the trigonal distortion of the local spin-orbit interactions with the spatial disposition of the constituent ions, a finding that can be extrapolated to other systems where SOC plays a leading role. Finally, and of particular significance for the molecular magnetism community, the unique observation of the behavior of the tunnel splittings at different resonances with the magnitude of an applied transverse magnetic field unveils the applicability of the spin selection rules within the nature of QTM, including tunneling in odd-numbered resonances, definitively resolving longstanding questions in the field.**

The observation of QTM in SMMs in the early 1990s [1-3] touched off a wave of explorations into the fundamental aspects of nanomagnetism and has since borne a wealth of fertile data and impacted on an extraordinarily broad range of science. One of the most prominent findings arose from the theoretical revelation of quantum phase interference as a modulator of QTM [4-6] and the subsequent experimental confirmation for several molecular symmetries [8-12], which established the importance of subtle contributions introduced by second (and higher) order molecular anisotropic interactions in shaping

QTM behavior. It is in the kernel of this understanding where one finds profound insight into the relationship between the SOC symmetries and QTM, including the symmetry-imposed spin selection rules that must be satisfied in order to break the energy degeneracy and allow tunneling to occur between a pair of molecular spin levels, labeled as *m* and *m'*, at a QTM resonance *k*, defined as $k = m'-m$. It follows that only resonances where *k* is an integer multiple of the lower molecular symmetry order are allowed. As such, in molecules of rhombic symmetry, only resonances corresponding to a multiple of two are unfrozen, while trigonal and tetragonal symmetries only lift state degeneracies at resonances $k = 3 \times n$ and $k = 4 \times n$ (*n* = integer), respectively. These apparently clear restrictions associated with the SOC symmetry have puzzled researchers in the field for two decades, as quantum relaxation has been observed in all QTM resonances for most SMMs regardless of their respective molecular symmetry. The only exception so far has been a $Mn_3$ SMM of trigonal symmetry [13], in which the absence of a resonance ($k = 1$) provided the first clear evidence of spin selection rules in QTM. However, the observation of other resonances also forbidden by symmetry (i.e. $k = 2$) in that molecule and the impossibility of studying the detailed field behavior of the different tunnel splittings have dimmed the relevance of that finding, since its interpretation has relied exclusively upon theoretical analyses derived from indirect results (see [14-16]).

Interestingly, the lowest SOC symmetry that supports QTM in odd-numbered resonances is trigonal. This is an important and fundamental system to explore, since only a transverse magnetic field can also break the degeneracy between the spin levels at odd-numbered resonances. Nevertheless, odd resonances are commonly observed in experiments performed in the absence of transverse field, while internal fields (e.g.

dipolar or nuclear fields) are not sufficiently large to explain the observed tunneling rates. It is partly for this reason that the study of QTM in molecules of trigonal symmetry has become an important quest within the molecular magnetism community. Unfortunately, as common as it may be in other realms of nature – e.g. the trigonal disposition of leaves around the stem in a three-leaf clover, which acts to maximize the energy influx from sunlight – trigonal symmetry has remained an elusive target for QTM in molecular magnets. Out of the several hundred SMMs synthetized to date, only a few dozen present trigonal site symmetry, and most of those are formed by transition metal trimers that couple antiferromagnetically, resulting in spin frustration and weakly defined spin ground states. Of the few exhibiting ferromagnetic coupling, a large portion present other issues, such as significant inter-molecular interactions or the coexistence of different species (see e.g. Refs. [17,18]), which have prevented the appropriate study of the QTM under this particular symmetry. Indeed, the only direct physical manifestation of a trigonal molecular symmetry has been recently reported by Sorace et al. [16] in a heteronuclear $Fe_3Cr$ SMM, for which the electron paramagnetic resonance (EPR) spectra shows a six-fold modulation as a function of the angle of application of a transverse magnetic field within the hard anisotropy plane of the molecule. However, according to the authors, extremely fast tunneling rates prevented the desired study of QTM in that compound.

In this Report we present the first manifestation of a three-fold modulation of the QTM rates for a SMM of trigonal symmetry, as well as a number of related fascinating behaviors, including the first observation of a spatial corrugation of the SOC energy landscape, which represent an important step forward in the effort to reconcile the theory of QTM with observation, and which sheds light onto the answers of many longstanding questions.

The SMM complex we studied has chemical formula $Mn_3O(Et\text{-}sao)_3(Et\text{-}py)_3ClO_4$ (henceforth referred to as "$Mn_3$"). Chemical analysis [17] ascribes the magnetic behavior to a metallic core containing three $Mn^{III}$ ions ($s = 2$) ferromagnetically coupled via a superexchange interaction that acts across O bridges, resulting in a $S = 6$ ground state at low temperature. A schematic constructed from X-ray diffraction measurements is inset to Fig 1 and details the magnetic core (see Ref. [17] for more details).

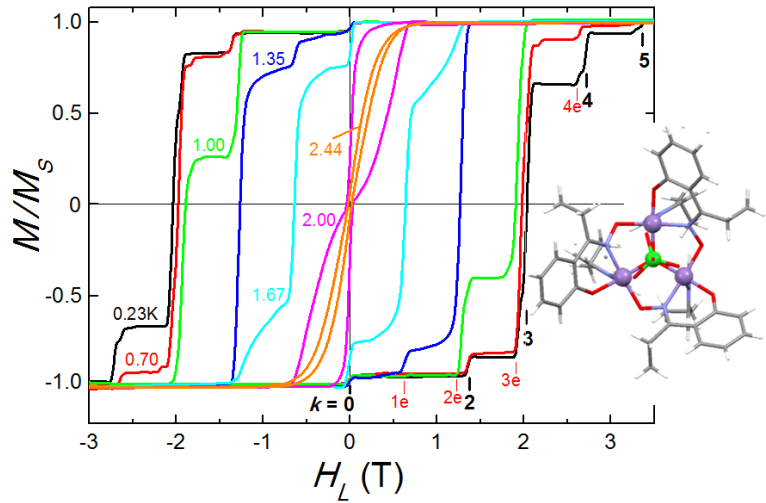

*Fig. 1:* *Stepwise magnetic hysteresis loops characteristic of resonant QTM obtained in a single crystal of $Mn_3O(Et\text{-}sao)_3(Et\text{-}py)_3ClO_4$ SMMs at different temperatures. Up to six resonances can be observed ($k = 0, \pm1, \pm2, \pm3, \pm4$ and $\pm5$), including steps associated with QTM through excited states ($k = 1e, 2e, 3e,...$). The inset shows the $Mn_3$ core: Mn (purple), Cl (green), O (red), N (lavender), C (grey) and H (white).*

Fig. 1 shows magnetization hysteresis loops obtained from a single crystal of Mn$_3$ SMMs with the field applied along the easy anisotropy axis (*z*-axis) at different temperatures. The sharpness of the observed QTM resonances, labeled $k = 0–5$, indicates the high quality of the crystal, typical of SMMs crystallized without solvent molecules. Note that resonance $k = 1$ is absent at temperatures below 1.35 K. First observed in a Mn$_3$ SMM [13], this effect is a consequence of the SOC selection rules discussed above, which make this resonance forbidden under trigonal symmetry considerations (i.e. $k \neq 3 \times n$). The QTM spectroscopy (i.e. position of the resonances) in this figure allows for the determination of the spin Hamiltonian governing the sample's quantum dynamics. A typical approach is to describe the molecule as a rigid spin (*S*) composed of the interacting single-ion spins ($s_i$), which is known as the giant spin approximation (GSA). This description is constructed out of an interaction Hamiltonian which includes terms representing intrinsic spin-orbit anisotropic interactions and the Zeeman coupling between the giant spin and an applied magnetic field, which for trigonal symmetry can be written as follows:

$$\hat{H}_{GSA} = D\hat{S}_z^2 + B_4^0 O_4^0 + B_4^3 O_4^3 + B_6^6 O_6^6 + \mu_B \boldsymbol{B} \cdot \vec{g} \cdot \hat{\boldsymbol{S}} \qquad (1)$$

The first four terms characterize the zero-field splitting (*zfs*) anisotropy, with the first usually dominant and responsible for the easy magnetization axis of the molecule (with a quartic axial correction given by the second). The Stevens spin operators ($O_p^q$) are restricted by the spin value ($p \leq 2S$) and the rotational symmetry, represented by $q$ ($\leq p$). Here we consider only second ($\hat{S}_z^2 = O_2^0$, with $D = 3B_2^0$) and fourth-order ($O_4^0 = \hat{S}_z^4$) axial terms and the leading trigonal ($O_4^3 = [\hat{S}_z, \hat{S}_+^3 \pm \hat{S}_-^3]$) and hexagonal ($O_6^6 = S_+^6 \pm \hat{S}_-^6$) transverse operators. The final term is the spin-field Zeeman interaction, parameterized by the Lande $g$-tensor, $\vec{\vec{g}}$. The positions of the QTM resonances in Fig. 1 can be well explained by diagonalization of the GSA Hamiltonian using an isotropic $g = 2$, $D = 0.86$ K and $B_4^0 = 1.4$ mK (the transverse anisotropy terms have a negligible effect on the spin projection energies, being only significant at degeneracies). Fig. S1 in *Supplementary Materials* shows the correspondence between the QTM spectroscopy data and the levels of the $S = 6$ spin multiplet.

In order to explore the nature of the QTM phenomena in this molecule, we will focus the following discussion on the behavior of the QTM resonances in the presence of a transverse magnetic field ($H_T$). In particular, we will describe the modulation of the QTM probability at resonances $k = 0$–3 as a function of both the angle of application and the magnitude of $H_T$, from which information about the QTM symmetry (and thus the SOC) can be extracted. We define the QTM probability $P_k$ as the normalized change in magnetization that occurs when the longitudinal field $H_L$ is swept through a resonance. This probability is related to the "tunnel splitting" ($\Delta_k$) that breaks the degeneracy between opposite spin projection levels, as given by the Landau-Zener formula [20], $P_k = 1 - \exp[-\pi \Delta_k^2 n / 2 \upsilon_0 \delta]$, where $\upsilon_0 = g\mu_B(2S - k)$, $\delta$ is the field sweep rate and $n$ is the number of times resonance $k$ is crossed. To extract the angular dependence of $P_k$, a fixed transverse field is maintained at a given angle $\phi$ within the molecular $xy$-plane while the longitudinal field is swept across the resonance under study. The process is then repeated for different $\phi$ ranging from 0 to 360 degrees. In order to optimize the quality of the results, and overcome several technical limitations, different protocols of measurement had to be followed for each resonance, as explained in Section 2 of the *Supplementary Materials*.

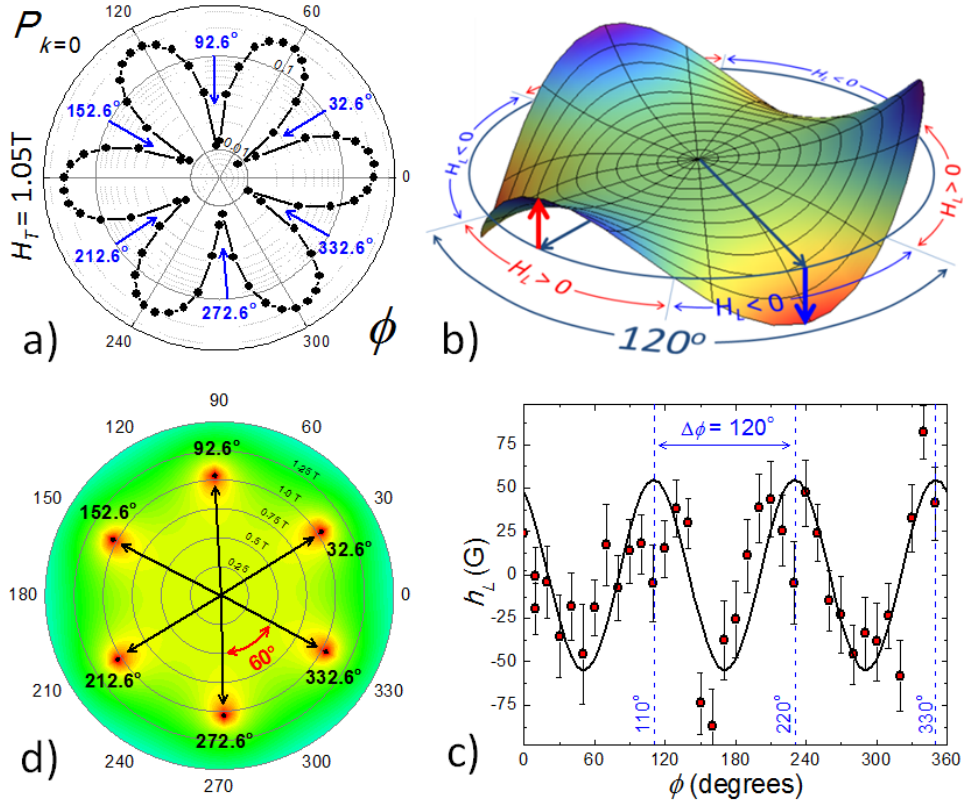

*Fig. 2:* (a) Six-fold angular modulation of the MQT probability as a function of the angle of application of a 1.05-T transverse field within the molecular xy-plane. Sharp minima appear spaced by 60 degrees and starting at 32.6 degrees. (b) Sketch illustrating the three-fold corrugation of the hard anisotropy plane of the $Mn_3$ SMM, which defines the longitudinal compensating field required to keep the system in resonance when a transverse field is applied. (c) Three-fold modulation of the compensating field measured in $Mn_3$ at 1.57 K (circles). The continuous line represents the fitting from diagonalization of the MS Hamiltonian in Eqn. (2). (d) Contour polar plot of $P_{k=0}$ vs $\phi$ and $H_T$, where the six-fold modulation of the BPI minima is shown to coincide with the observations in (a).

Let us focus first on resonance $k = 0$. Fig. 2a shows a polar plot of $P_{k=0}$ vs. $\phi$ where an extraordinary six-fold modulation emerges, with sharp minima occurring at angles $\phi_{\min,k=0}^{BPI} = 32.6^o + m\times 60^o$ which correspond to quenching of the tunnel splittings as a result of a destructive quantum interference effect, also known as Berry phase interference (BPI). A transverse field value of $H_T = 1.05$ T was purposely chosen to emphasize the BPI effect. However, this six-fold appearance can be misleading – the expected symmetry of the molecule is three fold, and so the shape of the resonance behavior should be as well (in fact we observe such modulation in all the other resonances, as discussed below). In the GSA, this apparent anomaly is a consequence of the trigonal transverse SOC anisotropy term, $O_4^3 = [\hat{S}_z, \hat{S}_+^3 \pm \hat{S}_-^3]$, which results from a commutation between the axial ($\hat{S}_z$) and the third-order creation and annihilation ($\hat{S}_+^3 \pm \hat{S}_-^3$) spin operators. Apart from generating a three-fold modulation of the anisotropy barrier (see Fig S2b), this term acts as an effective inherent longitudinal field that produces a three-fold corrugation of the hard anisotropy plane of the molecule in the presence of a transverse field, as illustrated in Fig. 2b. As a result of this corrugation, a "compensating longitudinal field" ($h_L$) must be added when a transverse field is applied in order to bring the system into resonance. The value of this compensating field oscillates between opposite polarities as dictated by the commutation with the third-order spin operators, forming a three-fold pattern of alternating sign in pace with, and thus obscuring, the three-fold modulation of the tunnel splitting in this resonance. This stands in contrast to the behavior of resonances where the longitudinal field required has a fixed value ($k > 0$) and the three-fold modulation can be clearly observed, as shown below. This is an extremely subtle effect and difficult to observe for the ground state splitting at resonance $k = 0$ (i.e. mixing states $m = +6$ and $m' = -6$) since the magnitude of $h_L$ ($< 3$ G), in the range of $H_T$ explored in these experiments (<1.2 T), is much smaller than the effective field-width of the resonance (i.e. ~2000 G at $H_T = 1.2$ T). As explained in *Section 2* of the *Supplementary Materials*, a sophisticated measurement protocol was employed in order to discern the angular modulation of the compensating field, involving relaxation measurements at high temperature ($T = 1.57$ K), at which QTM in $k = 0$ occurs predominantly through the third excited tunnel splitting (mixing states $m = +3$ and $m' = -3$). The corrugation of the hard anisotropy plane is much more pronounced in this splitting than in the lower ones as a result of its commensuration ($\Delta m = 3 \times n$) with the symmetry of the trigonal SOC term responsible for this effect, leading to much larger compensating field values. The results are displayed in Fig. 2c, where the compensating field shows an alternation between -55 and +55 Gauss with an overall three-fold oscillation pattern. Interestingly, its absolute maximum values, found at $\phi_{max}^{|h_L|} = 50° + n \times 60°$, do not coincide with the angular positions of the BPI minima in this resonance ($\phi_{min,k=0}^{BPI} = 32.6° + n \times 60°$), as would have been expected from the GSA Hamiltonian in Eqn. (1). Although an unphysical rotation of the $O_4^3$ and $O_6^6$ SOC terms about the $z$-axis in the GSA Hamiltonian can account for this effect, this shift can naturally be described by a set of specific tilts of the single-ion SOC tensors within a multi-spin (MS) Hamiltonian description, as it will be shown below.

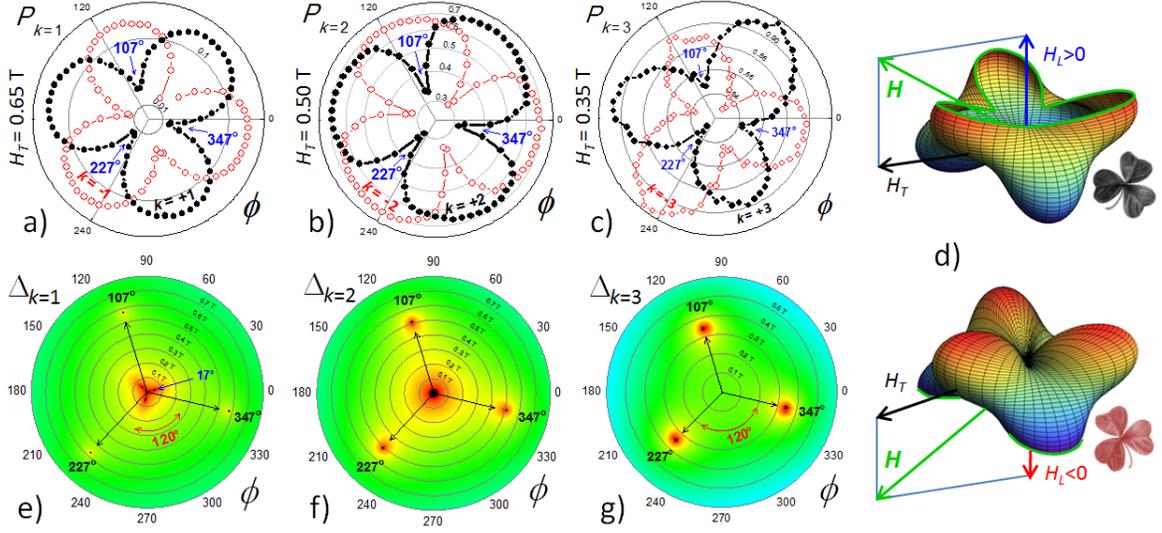

*Fig. 3:* (a-c) Polar plots of the QTM probability as a function of the angle of application of a transverse field $H_T = 0.65\,T$, $0.50\,T$ and $0.35\,T$, in resonances $k = 1, 2$ and $3$, respectively (solid black symbols). BPI minima are separated by 120 degrees (three-fold modulation) at angles $\phi_{\min,k>0}^{BPI} = 107^o + n \times 120^o$. The corresponding modulations for the opposite resonance polarities ($H_L < 0$) are presented with open red symbols. The latter depict a three-fold modulation of the BPI minima appearing at angles $\phi_{\min,k<0}^{BPI} = 47^o + n \times 120^o$, i.e. phase-shifted by 60 degrees with respect to the positive field resonances. (d) Sketches illustrating cuts by the total applied magnetic field vector into the anisotropy barrier generated by the trigonal transverse anisotropy term in the GSA Hamiltonian, which generate three-leaf clover forms that resemble the data in (a-c). (e-g) Contour polar plots of $P_{k=1}$, $P_{k=2}$ and $P_{k=3}$, respectively, vs $\phi$ and $H_T$, where the three-fold modulations of the BPI minima are found to coincide with the observations in (a-c).

The trigonal symmetry of this SMM becomes obvious in the resonances where a longitudinal magnetic field is applied, i.e. $k > 0$. Astounding three-fold angular modulations of the QTM probabilities are observed for resonances $k = +1, +2$ and $+3$ in Figs. 3a, 3b and 3c (solid black circles), respectively, with minima found at $\phi_{\min,k>0}^{BPI} = 107° + n \times 120°$, corresponding to conditions for BPI (again, the values of the transverse field have been purposely chosen to emphasize the modulation as much as possible). A fascinating consequence of this symmetry is that it produces anisotropy axes which are "hard" and "medium" at the same time, depending of the direction of application of both the longitudinal and transverse magnetic fields. Note that if the longitudinal field is reversed, as is the case of resonances $k = -1, -2$ and $-3$ in Figs. 3a, 3b and 3c (open red circles), respectively, the three-fold modulation is shifted by 60 degrees, with minima appearing at $\phi_{\min,k<0}^{BPI} = 47° + n \times 120°$. This is a feature of the time-reversal invariance of the SOC upon full reversal of the total magnetic field. To aid in understanding, cuts of the $O_4^3$-generated anisotropy barrier by the total applied field vector, $\boldsymbol{H}$, are illustrated in Fig. 3d for both polarities of the applied longitudinal field (i.e. opposite $k$-signs), resulting in three-leaf clover style shapes rotated by 60 degrees relative to each other, as observed in Figs. 3a-c.

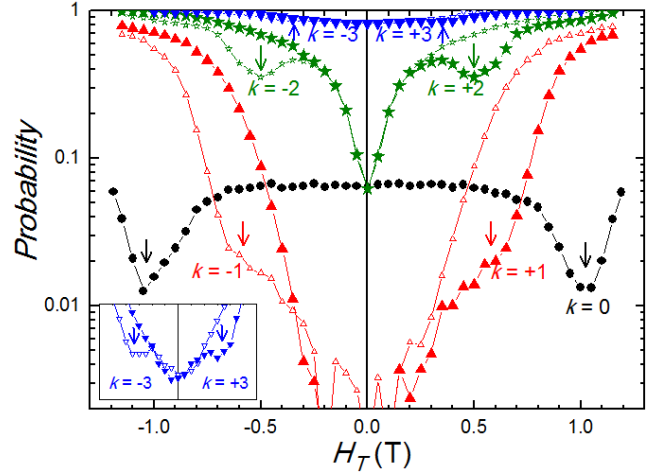

*Fig. 4:* *QTM probability of resonances |k| = 0–3 as a function of a transverse field applied along the following axes $\phi = 32.2°(+180°)$ for $k = 0$, and $\phi = 107°(+180°)$ for $|k| > 0$. Clear BPI minima are observed at $H_T = \pm1.05$ T, $\pm0.57$ T, $\pm0.50$ T and $\pm0.35$ T for resonances $k = 0, \pm1, \pm2$ and $\pm3$, respectively, as marked by the corresponding arrows. The inset shows a -0.6 to +0.6 T transverse field zoom of the $k = \pm3$ data in order to help locate the BPI minima for this resonance. Reversal of the longitudinal field produces the specular image with respect to reversal of the transverse field, as imposed by the time-reversal invariance of the spin-orbit interaction.*

We now turn our attention to the modulation of the QTM probabilities by the magnitude of a transverse magnetic field applied along the characteristic "hard/medium" directions within the molecular *xy*-plane, i.e. $\phi^{BPI}_{\min,k=0} = 32.6°(+180°)$ for $k = 0$ and $\phi^{BPI}_{\min,|k|>0} = 107°(+180°)$ for $k > 0$. The results are shown in Fig. 4 for all resonances; $k = 0$ (solid black circles), $k > 0$ (solid red, green, and blue data points) and $k < 0$ (open data points). BPI minima are found near $H^{BPI}_{T,k=0} = \pm1.05$ T, $H^{BPI}_{T,k=\pm1} = \pm0.57$ T, $H^{BPI}_{T,k=\pm2} = \pm0.50$ T and $H^{BPI}_{T,k=\pm3} = \pm0.35$ T (marked by arrows in Fig. 4). These are the same transverse fields chosen for the angular modulation measurements in Figs. 2 and 3 (with the exception of $k = \pm 1$, in which a value of 0.65 T was used instead). The GSA Hamiltonian in Eqn. (1) cannot account neither for the position of the BPI minima in all the resonances in Fig. 4 nor the difference in angles at which the BPI minima appear between resonances $k = 0$ (i.e. 32.6 + n×60°, Fig. 2a) and $|k| > 0$ (47° + n×60°, alternating with the sign of $k$, Fig. 3), amounting to a relative shift of $\Delta\phi = 14.4$ degrees. As mentioned above, a similar shift is also observed between the $k = 0$ BPI minima and the angular values of the compensating field maxima (~50° + n×60°, Fig. 2c), which also eludes an explanation with Eqn. (1). Interestingly, a 15-degree rotation of trigonal $O_4^3$ and hexagonal $O_6^6$ transverse anisotropy terms in Eqn. (1) about the $z$-axis can accurately account for all the observations (using $B_4^3 = -1.43 \times 10^{-4}$ K and $B_6^6 = 5.73 \times 10^{-7}$ K), as shown in Fig. S4. However, this represents an uninformed phenomenological attempt lacking physical foundation. A more natural approach, with real physical significance, is to employ a MS interaction Hamiltonian which takes into account the constituent ions and the corresponding intra-molecular interactions, as follows:

$$\hat{H}_{MS} = \sum_i \hat{s}_i \cdot \vec{R}_i^T \cdot \vec{d}_i \cdot \vec{R}_i \cdot \hat{s}_i + \sum_i g\mu_B \hat{s}_i \cdot \vec{B} + \sum_{i>j} \hat{s}_i \cdot \vec{J}_{i,j} \cdot \hat{s}_j \qquad (2)$$

where $\hat{s}_i$ is the spin operator of the $i$th ion, $\vec{d}_i$ is a diagonal 3x3 matrix with values $e_i$, $-e_i$ and $d_i$ (representing the rhombic and axial anisotropy terms of the $i$-th ion), and $\vec{J}_{i,j}$ is the exchange coupling tensor between each pair ($i,j$) of spins. This model not only permits consideration of the couplings between the spins of the constituent ions (therefore explaining the presence of excited spin multiplets and completely accounting for all the observed QTM steps, see Fig. S1), but it also allows for an arbitrary rotation of the single-ion SOC tensors, which is achieved by the Euler matrix $\vec{\vec{R}}_i^T$ and characterized by the Euler angles $\alpha_i$, $\beta_i$ and $\gamma_i$, as illustrated in Fig. 5a.

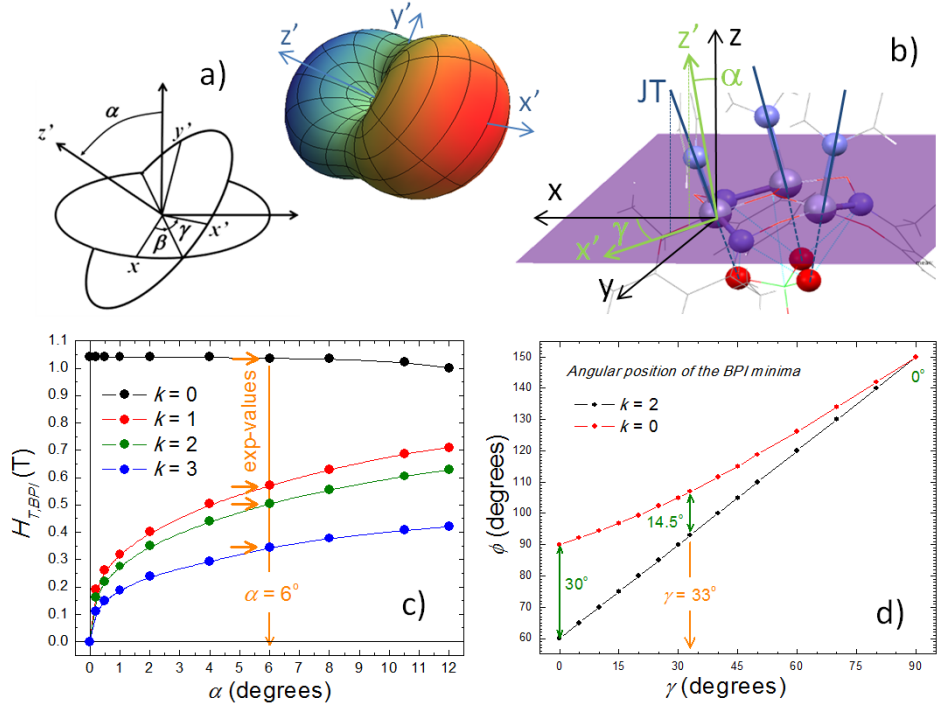

***Fig. 5:*** *(a) Arbitrary $\alpha$-$\beta$-$\gamma$ Euler rotation of the second-order SOC tensor of a single manganese ion. (b) Localization of the rotated easy (z') and "hard/medium" (x) anisotropy axes of a single Mn ion within the molecular arrangement of the 6-coordinate Mn-O (×4) and Mn-N (×2) bonds, and with respect to the direction of the Jahn-Teller axis, which lies along the N-Mn-O bonds. (c) Transverse field positions of the BPI minima of resonances k = 0–3 as a function of the Euler angle $\alpha$, which represents the tilt of the ion easy axis away from the overall molecular easy z-axis, which is perpendicular to the $Mn_3$ plane (purple plane in (b)). The arrows indicate the values observed experimentally (see Fig. 4), which are accounted for by an angle of $\alpha = 6^o$. (d) Calculated angular positions of the BPI minima in resonances k = 0 and k = 2 as a function of the rotation Euler angle $\gamma$. The observed angular shift of $\Delta\phi_{exp} = 14.5^o$ is theoretically matched with a value of $\gamma = 33^o$.*

Opportunely, $\alpha_i$ and $\gamma_i$ are equal for all ions (i.e. $\alpha$ and $\gamma$) and unambiguously determined by the particular behavior of the BPI minima patterns within the transverse field magnitude-angle phase space (demonstrating the importance of observing the BPI in this molecule), while $\beta_1 = 0º$, $\beta_2 = 120º$ and $\beta_3 = 240º$ are imposed by the trigonal symmetry. On the one hand, our simulations indicate that a tilting by an angle $\alpha$ of the easy anisotropy axis ($z'$ in Fig. 5a) away from the perpendicular to the molecular plane has a strong effect on the magnitudes of transverse field at which the minima occur for resonances $k = 1, 2, 3$. This dependence is shown in Fig. 5c as obtained from diagonalization of the MS Hamiltonian in Eqn. (2) with the following parameters: $g_i = 2$, $d = 3.6$ K, $e = 0.62$ K and isotropic $J = 3.1$ K (the fact that $J \sim d$ justifies the presence of low-lying spin multiplets). Note that $k = 0$ remains unaffected for small values of $\alpha$, which is not a surprise as this resonance is the only one allowed in the absence of any local ion tilts (the spin selection rule of hexagonal symmetry is $k = 6 \times n$). The positions at which we experimentally observe the minima are indicated in Fig. 5c, and coincide with the predicted values for a tilt of $\alpha = 6º$. On the other hand, the value of $\gamma$ (a rotation about the tilted easy $z'$- axis, see Fig. 5a) generates a phase shift ($\Delta\phi$) between the angular modulation of the BPI minima in $k = 0$ and that in the other resonances. Fig. 5d shows the calculated values of the transverse field vector angle at which the berry phase minima occur for the $k = 0$ and $k = 2$ resonances as a function of $\gamma$. For $\gamma = 0$, the 30-degree relative phase in the molecular frame angle $\phi$ between the two sets of minima modulation places the minima in the $k = 2$ modulation at angles where the $k = 0$ displays a maxima, whereas for $\gamma = 90$ degrees the minima are coincident in $\phi$. The experimentally observed difference between the phases of the $k = 0$ and $k > 0$ minima is $\Delta\phi_{exp} = 14.5$ degrees, which agrees with the calculated difference for an angle of $\gamma = 33°$ ($\Delta\phi_{th} = 14.4°$), as marked in Fig. 5d. This set of Euler rotation angles explains all the novel experimental findings provided in this Report, with the relevant simulations of the BPI behavior displayed in Fig. 2d and Figs. 3e-g, including the fitting of the compensating longitudinal field in Fig. 2c (see Supplementary Materials for details on this fitting), and data in Fig. S4, which shows the calculated behavior of the tunnel splittings for all resonances as a function of the transverse field magnitude, in direct correspondence to the experimental results given in Fig. 4. Another important aspect in these results is the substantially different transverse field dependence of the splittings in forbidden resonances $k = 1$ and $k = 2$, which is observed here for the first time. This difference, with $\Delta_{k=1}$ growing much more slowly than $\Delta_{k=2}$ with increasing transverse field (clearly observed in Fig. 4), is crucial in understanding the appearance of one of the two forbidden resonances, since the contribution of small internal transverse fields (dipole or hyperfine fields) can only unfreeze QTM in resonance $k = 2$ in the absence of an applied transverse field, while much larger field values would be necessary to similarly affect resonance $k = 1$. This is a potentially important conclusion because, together with the effect of local disorder-induced distortions (as discussed in Ref. [15]), it may explain why QTM is observed at all resonances in most SMMs regardless of the selection rules imposed by the SOC symmetry.

Given the precision with which we are able to associate theoretical parameters of the ions' orientations with the observed phenomena, we finally raise the tantalizing prospect of associating the measured SOC anisotropies with the specifics of the chemical arrangement. Based on the orientation of our crystal sample and identification of the various crystalline axes we can ascertain the most likely mapping, which ultimately hinges upon the degree of accuracy in determining the orientation of the sample (of an approximately flat triangular shape) within our apparatus. From X-ray face-indexing data of a single crystal, we can associate the plane formed by the three manganese ions (purple plane in Fig. 5b) with the flat face of the triangular crystal. A further association of the crystallographic [1,0,0] vector (defining one of the side-planes of the crystal) with the molecular $x$-axis allows the determination of the orientation of the rotated anisotropy axes of each ion within the molecular structure. Accordingly in this frame, the easy anisotropy $z'$-axis of the ions is tilted by an angle $\alpha = 6°$ from the perpendicular ($z$-axis) to the Mn$_3$ plane, and by an angle $\gamma = 33°$ from the $x'$ axis, as shown in Fig. 5b. In fact, this rotation places the easy anisotropy axis of the ions ~12 degrees away from the approximate line formed by the O-Mn-N bonds, along which the Jahn-Teller axes of the manganese ions are, to a first approximation, expected to lie. This is because the Mn ion is 6-coordinate and thus contains four short and two long bonds, the latter defining the Jahn-Teller axis. Prior characterizations in several other Mn$^{III}$ compounds have identified the Jahn-Teller axis orientations as significantly different from those of the anisotropy easy axis [21]. In a crystal of the Mn$_3$ sample, there are actually two molecular species coexisting within the unit lattice cell. Luckily, although different, their respective orientations are such that both species behave equally upon the application of a magnetic field and hence are magnetically indistinguishable (as explained in Section 4 of the *Supplemental Materials*).

Our magnetization studies have shown a clear correlation between the chemical structure and the form of the SOC anisotropy/energy landscape of the spin of a SMM, and represent a nearly full treatment of QTM phenomenon. By illustrating the potential for such high-resolution examinations of the molecular symmetry, we see a vast and rich frontier remaining to be explored by the pairing of molecular engineering and low temperature physics experiment.

**Acknowledgements:** J.H.A. and E.d.B acknowledge support from the National Science Foundation (DMR#0747587). E.K.B. thanks EPSRC for funding.

## METHODS

**Magnetization Measurements** were carried out on a sub-millimeter sized single crystal placed on top of a high-sensitivity micro-Hall effect magnetometer. Magnetization hysteresis loops were recorded in the presence of magnetic fields generated by a superconducting vector magnet, capable of generating arbitrarily oriented vectors of field at the sample location with components reaching magnitudes of up to 8 T along one axis and up to 1.2 T along the two perpendiculars. The lowest stable temperature achievable in our Oxford Instruments $^3$He cryostat was 230 mK, which was the temperature reading for the majority of experiments conducted for this work (except data in Figs. 1 and 2c).

# SUPLEMENTARY MATERIALS

## Section 1: Giant Spin Approximation vs. Multi-Spin Hamiltonian

In this section, we discuss the differences and relations between the two theoretical descriptions used to explain the results presented in the main text: The giant spin approximation (GSA) and the multi-spin (MS) model.

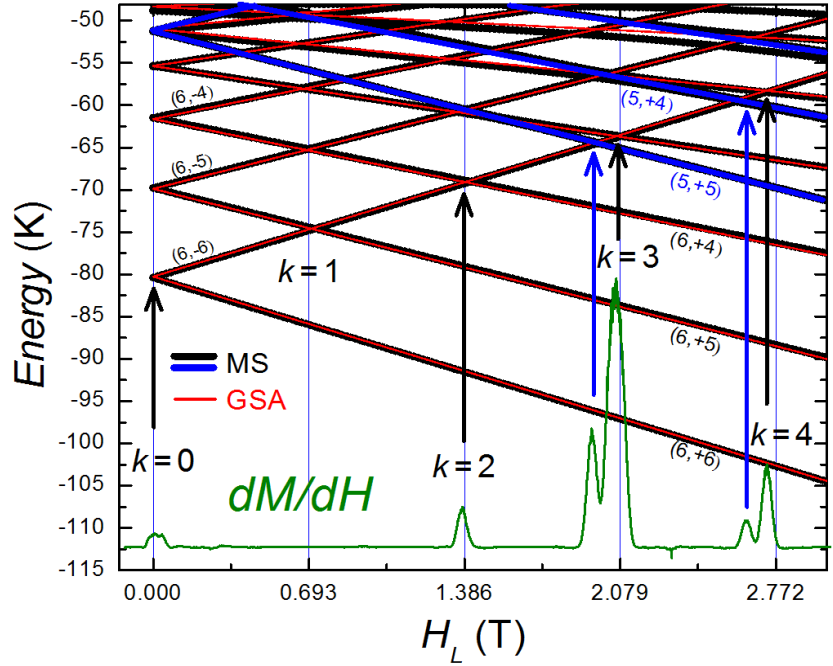

*Fig. S1:* *Energy levels extracted from the diagonalization of the GSA (red lines) and MS (black and blue lines) Hamiltonians with the parameters given in the main text. The arrows indicate the correspondence between the QTM resonances (level anti-crossings) and the observed QTM steps in the hysteresis loop obtained at the lowest temperature (= 230 mK), overlaid with a plot of its field derivative (green data). Resonance k = 1 is absent at this temperature due to the spin selection rules imposed by this symmetry.*

Fig. S1 displays the field derivative of the $T = 230$ mK hysteresis loop in Fig. 1 (green data) as well as the energy levels generated by diagonalization of both the GSA and the MS Hamiltonians given in the main text by Eqns. (1) and (2), respectively. The levels recreated by the GSA, corresponding to a solid spin $S = 6$ (red lines), can account for the main QTM resonances observed at low temperature (marked by black arrows in Fig. S1). However, the GSA cannot account for level crossings with excited spin multiplets (marked with blue arrows in Fig. S1). These can only be reproduced by the MS Hamiltonian (black and blue lines in Fig. S1), which takes into account the interaction between the single-ion spins and allows for the presence of low-lying excited multiplets, e.g. $S = 5$ (blue lines).

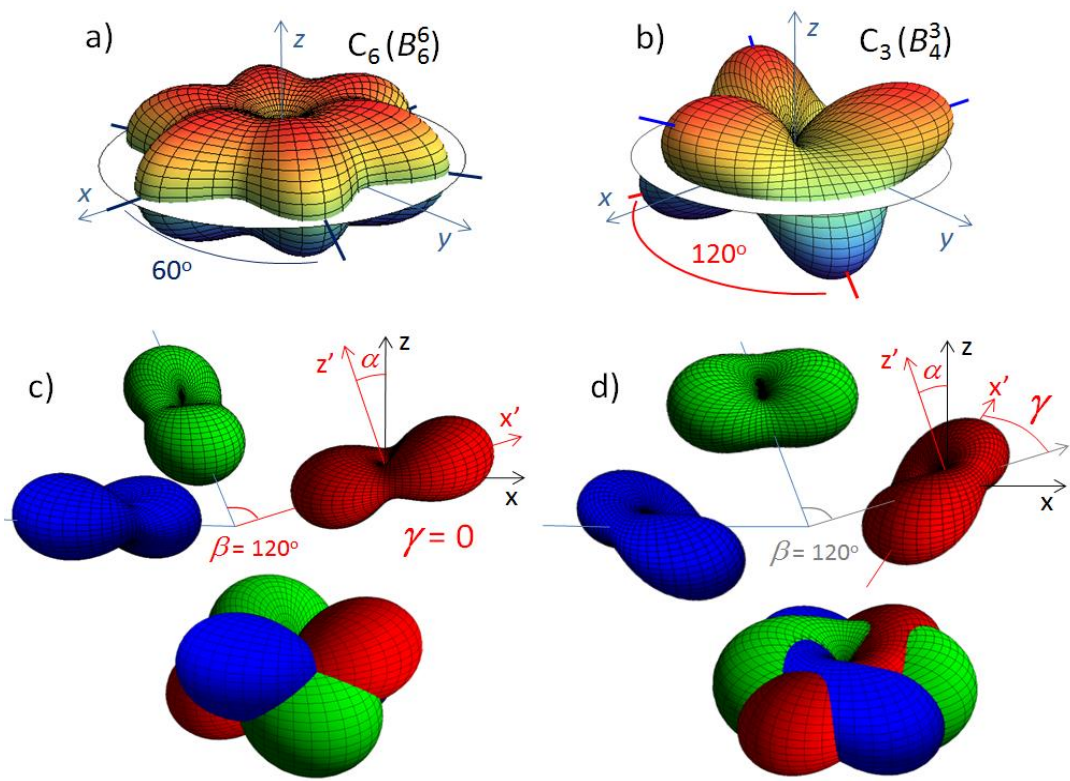

*Fig. S2: (a) and (b) show the anisotropy barriers generated by the hexagonal (a) and trigonal (b) transverse anisotropy terms included in the GSA Hamiltonian of Eqn. (1). (c) and (d) illustrate the anisotropy barriers resulting from the superposition of the single-ion second-order barriers for different Euler rotations, (a) $\gamma = 0$ and (b) $\gamma > 0$.*

Figure S2a and S2b show the anisotropy barriers generated by the hexagonal and trigonal transverse anisotropy terms of the GSA Hamiltonian in Eqn. (1), which are the lowest order transverse terms allowed by the trigonal molecular symmetry. As discussed in Refs. [14,15], the emergence of these two terms can be traced down to the symmetries imposed by the orientations of the single-ion anisotropy tensors of the three constituent ions. Assuming that the intrinsic symmetry of the *zfs* tensors of the individual ions is only 2nd order, the Hamiltonian of a single $Mn^{III}$ ion possesses $D_{2h}$ symmetry, characterized by three mutually orthogonal $C_2$ axes, coinciding with the three main anisotropy axes – namely, easy (local z-axis), hard (local x-axis) and medium (local y-axis). In the case that the Jahn-Teller axes of all the manganese ions are parallel ($\alpha = 0$, not shown in Fig. S2), the local z-axis of each $Mn^{III}$ coincides with the molecular $C_3$ axis. The resultant Hamiltonian should then possess $C_3 \times C_2 \times C_i = C_{6h}$ symmetry at zero longitudinal field ($C_6$ in the presence of a magnetic field). Note that the inversion operation $C_i$ results from the time-reversal symmetry of the spin-orbit interaction. According to this, the trigonal transverse term in the Hamiltonian must be zero ($B_4^3 = 0$) and the lowest allowed transverse term is hexagonal ($B_6^6 \neq 0$). In this case, the spin anisotropy barrier generated by the GSA Hamiltonian shows three hard and three medium axes leading to a six-fold modulation of the energy within the hard anisotropy *xy*-plane, with maxima separated by 60 degrees, as shown in Fig. S2a, with the *xy*-plane as a mirror symmetry plane. In the case that the Jahn-Teller axes are tilted away from the molecular *z*-axis ($\alpha > 0$, see Fig. S2c), the $C_2$ ion axes and $C_3$ molecular axes do not coincide, breaking the *xy*-mirror symmetry of the molecule and lowering the rotational symmetry to three-fold, i.e. $C_3 \times C_i = S_6$, which reduces to $C_3$ in the presence of a longitudinal field. This allows the inclusion of a trigonal transverse term in the GSA Hamiltonian ($B_4^3 \neq 0$), generating the anisotropy barrier depicted in Fig. S2b. It is easy to see that in the presence of a magnetic field applied along the *z*-axis (longitudinal field) the angular modulation of the energy is three-fold, with maxima separated by 120 degrees (as shown in Fig. S2b). Reversal of the longitudinal field results in a symmetric trigonal modulation with a pattern of maxima shifted by 60 degrees with respect to the original. In general, the results should be invariant with respect to reversal of the total magnetic field, as guaranteed by the time-reversal symmetry of the spin-orbit interaction.

One can easily see the correspondence between the anisotropy barriers generated by the trigonal transverse SOC term $O_4^3$ under the GSA (Fig. S2b) and by the simple superposition of the single-ion second-order anisotropies when the Jahn-Teller axes are tilted by arbitrary Euler angles α and β, while keeping γ = 0 (Fig. S2c), since they result in a similar spatial arrangement. However, the visual correspondence is not that simple if the single-ion SOC tensors are also tilted by an arbitrary angle γ > 0 (Fig. S2d), as required to interpret the results in this work. The reproduction of the resulting anisotropy barrier is not possible with a simple addition of the trigonal and hexagonal terms in the GSA Hamiltonian. A physical rotation about the z-axis of one term with respect to the other is required, which also accounts for the different modulations of the tunnel splittings, as discussed in *Section 3*.

## Section 2: Measurement of the QTM probability

Here we describe the exact protocols followed to study the QTM in each resonance in the presence of a transverse field. As mentioned in the Methods section, a single crystal of $Mn_3$ SMMs is placed on top of a 2DEG-based Hall-effect magnetometer, from which the Hall voltage is used to extract the magnetization state of the sample. Since these experiments require large transverse fields (<1.2 T) which need to be applied away from the sensor plane, the quantum Hall effect (occurring at fields over ~0.3 T perpendicular to the sensor plane) cannot be avoided. This, in addition to the presence of thermal avalanches producing abrupt changes in magnetization and nearby excited states in some resonances (e.g. $k = 3$), necessitated the development of a clever method to measure the QTM steps of the hysteresis loops that allowed us to elude these technical issues, which we describe below.

*Remainder method*: The basic principle behind this method is that one can measure some portion of hysteresis-style data by first measuring the whole, then everything *except* that portion, and then subtracting the two. This conveniently allowed us to circumvent conditions (i.e. Quantum Hall effect) that rendered our measurement device inoperable. Our method, illustrated in Fig. S3, began with the measurement of a "Base" loop (in the absence of any $H_T$) from which the total magnitude of the signal between opposite polarities of sample saturation could be determined. The next step was typically to sweep to a longitudinal field just before the resonance to be measured under conditions identical to the "Base", at which point the desired $H_T$ vector was generated. The longitudinal field would then be swept through the resonance, and then the transverse field turned off. The longitudinal field could then be swept to saturation, and recorded as the "Remainder" to be subtracted from the "Base", allowing determination of the probability of the resonance that took place while the transverse field was applied.

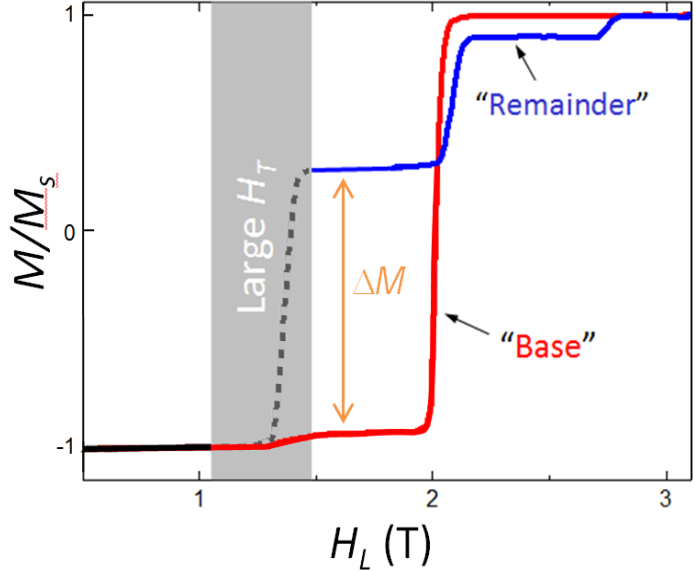

*Fig. S3: Schematic detailing the remainder method, which permitted indirect measurement of the QTM probability at a resonance. In this plot, the data ranges from negative to positive sample saturation. The resonance being measured (in the grayed region) would occur while our Hall bar device was affected by the Quantum Hall phenomenon, which temporarily degraded the integrity of the sensor. By measuring the total signal size with a "Base" measurement (red line) and the fraction of the sample that did not relax (the "Remainder", blue) in the absence of any transverse field, we were able to determine the fraction that relaxed during the resonance being measured.*

Due to the peculiarities of every resonance, different implementations of this method had to be used for different resonances. Below we provide a list with the details.

1. <u>Measurement of $P_{k=0}$ vs. $\phi$ at resonance $k = 0$</u>

    Here the longitudinal field was swept to negative saturation, then to a small negative field just before the resonance, at which point a transverse field of magnitude 1.05 T was applied and the longitudinal field swept through the resonance at a rate of 0.04 T/m. The sweep was then reversed such that the resonance was traversed again, which was performed in order to amplify the relatively small measured probability of the resonance. Once this second pass was completed, the transverse field was turned off, and the field swept back to negative saturation while the remainder was recorded. This was repeated for

different angular orientations of the transverse field vector, forming a complete rotation in 5 degree increments.

2. <u>Measurement of the compensating field at resonance $k = 0$</u>

   To measure the shift in the longitudinal field position of the k=0 resonance (i.e. the compensating field), a variation of the method described above had to be employed. As mentioned in the main text, these measurements were performed at 1.57 K in order to obtain QTM relaxation through the third excited tunnel splitting in this resonance. For initializing the measurement, a large negative longitudinal field was applied in the absence of a transverse field and then swept back to a value close to but far enough from the resonance to avoid relaxation, Then both the transverse and the longitudinal fields were swept at rates of 0.05 and 0.5 T/min, respectively, to the following values: 1.2 T (transverse field) and ~1000 G before the center of the resonance (longitudinal field). The resonance width is ~2000 G for HT = 1.2 T, and so QTM relaxation occurs during the process. Immediately after reaching those conditions, both fields were swept back (at the same rates) to the original point, after which the longitudinal field was swept in the absence of a transverse field to measure the remainder. The process was repeated at different angles of application of the transverse field within the xy-plane from $\phi = 0$ to $\phi = 360°$. Note that during the time that both fields are swept into the resonance, the tunneling conditions vary since the transverse field (a strong modulator of the tunnel splitting) varies continuously from zero to 1.2 T. However, given the fast relaxation at this temperature, a constant transverse field of 1.2 T could not be employed for this measurement. As we show below, the exponential dependence of the tunnel splitting on the transverse field implies the average tunneling rate is dominated by the largest values of the transverse field.

From these measurements we obtained a three-fold angular modulation of the tunnel probability due to the shift of the resonance caused by the alternating compensating field, as explained in the main text. In other words, the probability is higher when the final longitudinal field is closer to the resonance, and vice versa. Measurements following the same protocol but sweeping the longitudinal field to different points relative to the resonance allowed the conversion of the probability into a longitudinal field value (i.e. the compensating field data depicted in Fig. 2c).

For the fitting of the observed modulation of the compensating field (continuous line in Fig. 2c), values were extracted for different transverse fields (0.7-1.2 T) from diagonalization of Eqn. (2) with the parameters given in the main text. Note that the value of the compensating field grows with the transverse field. The different values were correspondingly weighted according to the probability calculated from the Landau-Zener formula in order to take into account the varying transverse field during the measuring process. As mentioned above, the MQT in this process is dominated by the largest transverse field values, with negligible effects for values below 0.7 T. The results are in excellent agreement with the observations.

3. Measurement of $P_{k=0}$ vs. $\phi$ at resonance $k = 1$

Here the longitudinal field was swept to (negative for $k = +1$, positive for $k = -1$) saturation, then to a value a small distance before the resonance, at which point a transverse field of magnitude 0.65 T was applied and the longitudinal field swept through the resonance at a rate of 0.1 T/m. Once this was completed, the transverse field was turned off and the field swept to (positive for $k = +1$, negative for $k = -1$) saturation while the remainder was recorded. This was repeated for different angular orientations of the transverse field vector, forming a complete rotation in 5 degree increments.

4. <u>Measurement of $P_{k=0}$ vs. $\phi$ at resonance $k = 2$</u>

   The same exact procedure that for k=1, using a transverse field of 0.5 T and sweeping the longitudinal field through the resonance at 0.1 T/min.

5. <u>Measurement of $P_{k=0}$ vs. $\phi$ at resonance $k = 3$</u>

   Here the longitudinal field was swept to (negative for $k = +3$, positive for $k = -3$) saturation, then to a value a small distance before the $k = \pm 2$ resonance, at which point a transverse field of magnitude 0.4 T was applied at an angle of 35 degrees and the longitudinal field swept through that resonance at a rate of 0.075 T/m. This relaxed roughly 50% of the spins, preventing avalanching of the $k = 3$ resonance that would otherwise occur. Once this was completed, the transverse field was turned off and the field swept to a value a small distance before the $k = 3$ resonance, at which point a transverse field of magnitude 0.35 T was applied and the longitudinal field swept through the resonance at a rate of 0.05 T/m. Once this was completed, the transverse field was turned off and the field swept to (positive for $k = +3$, negative for $k = -3$) saturation while the remainder was recorded. This was repeated for different angular orientations of the transverse field vector, forming a complete rotation in 5 degree increments.

6. <u>Measurement of $P_{k=0}$ vs. $H_T$ at resonance $k = 0$</u>

    This measurement was identical to that described in *Step 1*, except that the angle was held fixed at 32.6 degrees (direction of one of the minima in Fig. 2a) and the magnitude of the transverse vector applied was increased in 0.05 T increments from -1.19 to 1.19 T.

7. <u>Measurement of $P_{k=0}$ vs. $H_T$ at resonance $k = 1$</u>

    This measurement was identical to that described in *Step 3*, except that the angle was held fixed at 107 degrees (direction of one of the minima in Fig. 3a) and the magnitude of the transverse vector applied was increased in 0.05 T increments from -1.15 to 1.15 T.

8. <u>Measurement of $P_{k=0}$ vs. $H_T$ at resonance $k = 2$</u>

    This measurement was identical to that described in *Step 4*, except that the angle was held fixed at 107 degrees and the magnitude of the transverse vector applied was increased in 0.05 T increments from -1.15 to 1.15 T.

9. <u>Measurement of $P_{k=0}$ vs. $H_T$ at resonance $k = 3$</u>

This measurement was identical to that described described in *Step 5*, except that the angle was held fixed at 107 degrees and the magnitude of the transverse vector applied was increased in 0.05 T increments from -1 to 1 T.

## Section 3: Calculated Transverse Field Dependence of the tunnel splittings

Fig. S4 shows the ground state tunnel splittings of resonances $k = 0, \pm1, \pm2$ and $\pm3$, obtained by diagonalization of the MS Hamiltonian in Eqn. (2) with the parameters given in the main text. The positions of the BPI minima are in excellent agreement with the values observed in the experiment (Fig. 4). Note that a reversal of the longitudinal field produces the specular result with respect to reversal of the transverse field, as expected from the time-reversal symmetry of the spin-orbit interaction for a full reversal of the total magnetic field. In order to obtain the same results with the GSA, a 15-degree rotation between the trigonal $O_4^3$ and hexagonal $O_6^6$ transverse anisotropy terms about the z-axis of the molecule is required. The results obtained with the parameters given in the main text are shown in Fig.S4 for the positive resonances (continuous thin lines). The mapping of the MS Hamiltonian into the GSA enables the accurate determination of the origin of the different SOC terms in the Hamiltonian to an extent never achieved before.

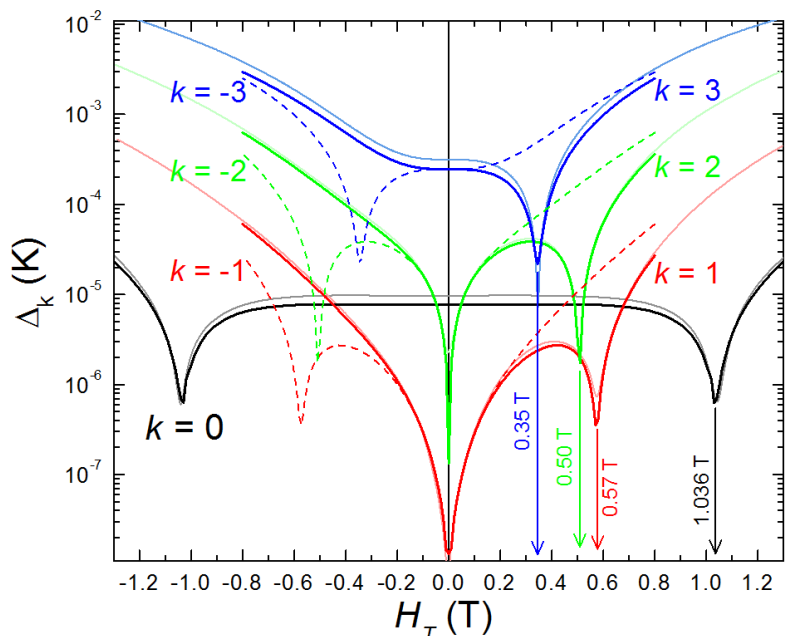

***Fig. S4:*** *Tunnel splittings of resonances |k| = 0–3 calculated from diagonalization of the MS Hamiltonian in Eqn. (2). The thin lines represent the results obtained using the GSA Hamiltonian of Eqn.(1) with a 15-degree rotation between the $O_4^3$ and $O_6^6$ transverse anisotropy terms.*

## Section 4: Structural details of the sample: Two coexisting molecular species

Details of the packing structure reveal that the crystal is built of equal numbers of two species, each inverted with respect to each other such that the Mn-Mn-Mn-Cl tetrahedrons point antiparallel and are rotated by 60 degrees about the perpendicular to the plane formed by the three manganese ions, a remarkable and fortunate symmetry in which molecules of both orientations respond indistinguishably to the experiments that are described in this Letter. Fig. S5 shows the two molecular species with the correspondence between respective equivalent ions and their respective main bonds (as indicated by the circles, squares and triangles). To illustrate the equivalency, the easy axes of the second-order anisotropy barriers of the ions are placed parallel to the Mn-$N_1$ bond, and different colors are used to identify different equivalencies. Note that, for example, the red SOC tensors in both species have the easy axis oriented in the same direction, and the same applies to the green and blue tensors. Therefore, these equivalent ions will behave equally under the action of a magnetic field. Since the exchange interaction is isotropic, this ion-ion equivalency between species will extend into the whole molecules, regardless of the species.

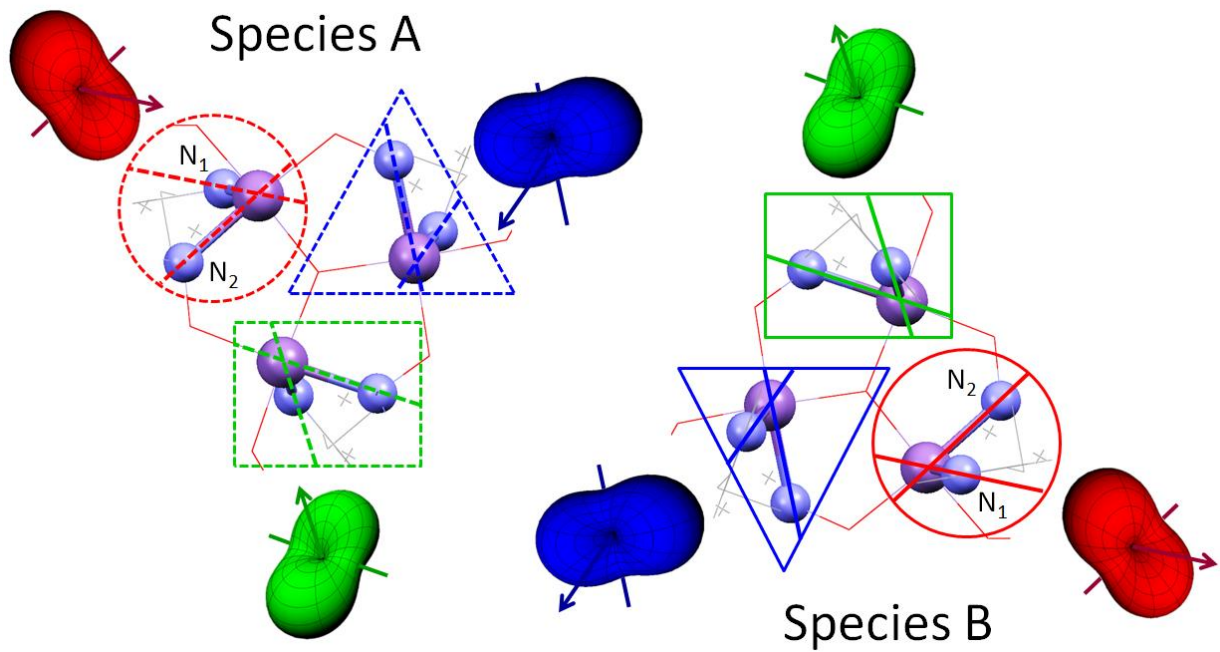

***Fig. S5:** The two molecular species in $Mn_3O(Et\text{-}sao)_3(Et\text{-}py)_3ClO_4$. Symbols and colors are used to identify the equivalencies between different ions in each species, which behave indistinguishable upon application of a magnetic field.*